\newcommand{\CLASSINPUTtoptextmargin}{0.8875in}
\newcommand{\CLASSINPUTbottomtextmargin}{1.1in}
\documentclass[conference, 10pt]{IEEEtran}
\IEEEoverridecommandlockouts
\usepackage{stmaryrd}
\usepackage{amsmath}
\usepackage{amsthm}
\usepackage{xfrac}
\usepackage{amssymb}
\usepackage{mathabx}
\usepackage{psfrag}
\usepackage{graphicx}
\usepackage{trimclip}
\usepackage{epstopdf}
\usepackage{cite}
\usepackage{algorithmic}
\usepackage{algorithm}
\usepackage{svg}
\usepackage{mathtools}
\usepackage{hyphenat}
\usepackage{booktabs}
\usepackage{accents}
\usepackage{hyperref}
\usepackage{multirow}
\usepackage{multicol}
\usepackage{accents}
\usepackage{minibox}
\usepackage{fancyhdr}
\usepackage{kantlipsum}
\usepackage{float}
\usepackage{multirow}
\usepackage{xparse,mathtools}
\fancypagestyle{fancy}

\usepackage{amsmath,amsfonts,amssymb}
\usepackage{algorithmic}
\usepackage{array}

\usepackage{cite}
\usepackage{amsmath,amssymb,amsfonts}
\usepackage{algorithmic}
\usepackage{graphicx}
\usepackage{textcomp}
\usepackage{xcolor}

\fancypagestyle{plain}{
  \fancyhf{}
  \fancyhead[C]{Conference on \LaTeX}     
  \fancyfoot[L]{This is a notice}

}
\usepackage{eso-pic}

\newlength{\dhatheight}

\def\BibTeX{{\rm B\kern-.05em{\sc i\kern-.025em b}\kern-.08em
    T\kern-.1667em\lower.7ex\hbox{E}\kern-.125emX}}

\usepackage{algorithm}
\usepackage{algorithmic}
\usepackage{amsthm}

\DeclareMathOperator{\diag}{diag}
\DeclareMathOperator{\myvec}{vec}
\DeclareMathOperator{\myspan}{span}
\DeclareMathOperator{\rank}{rank}

\newtheorem{lemma}{Lemma}

\newcommand{\mtrian}{\mathrel{\raisebox{-0.1ex}{%
\scalebox{0.8}[0.6]{$\vartriangle$}}}}

\let\underbrace\LaTeXunderbrace

\begin{document}

\AddToShipoutPictureBG*{%
  \AtPageUpperLeft{%
    \hspace*{\dimexpr0.175\paperwidth\relax}
    \minibox[c]{\\ \\ \\ \\ \emph{This work has been accepted to IEEE Military Communications Conference (MILCOM) 2023.} \\ \emph{Copyright may be transferred without notice, after which this version may no longer be accessible.}}
}}
\AddToShipoutPictureBG*{%
  \AtPageLowerLeft{%
    \setlength\unitlength{1in}%
    \hspace*{\dimexpr0.5\paperwidth\relax}
    \makebox(0,0.75)[c]%
}}

\title{Towards Explainable Machine Learning: The Effectiveness of Reservoir Computing in Wireless Receive Processing

\thanks{S. Jere, K. Said and L. Liu are with \emph{Wireless@VT}, Bradley Department of ECE, Virginia Tech. 
L. Zheng is with EECS Department, Massachusetts Institute of Technology.
This work was funded in part by U.S. National Science Foundation (NSF) under grants CCF-2003059 and CCF-2002908.}
}
\author{Shashank Jere, Karim Said, Lizhong Zheng and Lingjia Liu \\
}


\maketitle

\begin{abstract}
Deep learning has seen a rapid adoption in a variety of wireless communications applications, including at the physical layer. While it has delivered impressive performance in tasks such as channel equalization and receive processing/symbol detection, it leaves much to be desired when it comes to explaining this superior performance. In this work, we investigate the specific task of channel equalization by applying a popular learning-based technique known as Reservoir Computing (RC), which has shown superior performance compared to conventional methods and other learning-based approaches. 
Specifically, we apply the echo state network (ESN) as a channel equalizer and provide a first principles-based signal processing understanding of its operation. With this groundwork, we incorporate the available domain knowledge in the form of the statistics of the wireless channel directly into the weights of the ESN model. This paves the way for optimized initialization of the ESN model weights, which are traditionally untrained and randomly initialized. Finally, we show the improvement in receive processing/symbol detection performance with this optimized initialization through simulations. This is a first step towards explainable machine learning (XML) and assigning practical model interpretability that can be utilized together with the available domain knowledge to improve performance and enhance detection reliability. 

\end{abstract}


\section{Introduction}
\label{sec:Introduction}
The rise of deep learning in recent times has been unprecedented, owing largely to its remarkable success in a wide range of applications. The wireless communications field has also seen active adoption of machine learning (ML) and neural network (NN) based techniques at a rapid pace in a variety of problems and will play a significant role in next-generation wireless networks~\cite{Shafin2020}. 
The increasing complexity and modeling intractability of end-to-end wireless links caused by highly nonlinear radio frequency (RF) components and low-resolution analog-to-digital converters (ADCs) among others limits the applicability of traditional model-based approaches for most receive processing tasks such as channel equalization or receive symbol detection.
While state-of-the-art deep learning practice is that of training a large NN model ``offline'' with a large dataset and then deploying it for inference, this approach may not be feasible in wireless communications, especially at the physical layer where the over-the-air (OTA) training data is extremely limited.
Additionally, the choice of the NN model and its architecture may not be aligned with the nature of the specific problem at hand. 
On the other hand, reservoir computing (RC)~\cite{Lukosevicius2012} provides an ``online learning'' alternative whereby the model weights are adaptively updated through low-complexity training, making it ideal for application in tasks such as receive processing~\cite{zhou2020_LTD,zhou2020rcnet, RCstruct} and dynamic spectrum access~\cite{LiuAI2,Chang2020}, demonstrating superior performance in comparison to conventional model-based methods and other offline learning approaches~\cite{Khani2020,DSAComparison}. 
Despite this empirical evidence however, a systematic analysis of the general effectiveness of RC-based methods in physical layer receive processing operations is largely missing in state-of-the-art. 
Although there exist generalization error characterizations of RC from a statistical learning theory perspective~\cite{Gonon2020}, these do not address model interpretability or suggest how to incorporate domain knowledge, if available, into the NN design. 
Our recent work~\cite{Jere2023WCL} develops a first principles-based signal processing understanding of RC, specifically the echo state network (ESN), and provides basic interpretability to the conventionally untrained ESN model weights under the simple scenario of a two-tap fading channel. 
The primary contribution of this paper is a systematic analysis of the ESN as an equalizer in a general fading channel scenario with multiple taps, in addition to providing a clear procedure of incorporating available domain knowledge in the form of channel statistics directly into the ESN design.
We also assign interpretability to the conventionally untrained ESN model weights, thus representing a significant stride towards explainable machine learning in RC-based approaches when applied to receive processing tasks. 
\emph{Notation:} 
$\Re(\cdot)$ and $\Im(\cdot)$ are the real part and imaginary part operators respectively. $\mathbf{1}_N$ is the all-ones $N \times N$ matrix. 
$(\cdot)^*$ is the complex conjugate operator.
$\nabla(\cdot)$ is the gradient operator.

\section{System Model}
\subsection{Wireless Channel}
Consider a wireless channel with the discrete-time impulse response $\mathbf{h}=[h_0, h_1, \ldots, h_{L-1}]^T \in \mathbb{C}^{L} $. The system response can be written in terms of the $z$-transform as $ H_{\text{ch}}(z) = \sum_{\ell=0}^{L-1} h_{\ell}z^{-\ell}$.
Then, the frequency response of the channel, i.e., its Discrete-Time Fourier Transform (DTFT) is given by $ H_{\text{ch}}(j\omega) = \sum_{\ell=0}^{L-1} h_{\ell}e^{-j\ell \omega}$,
where $\omega$ is the digital frequency with $\omega \in [0,2\pi]$ rad/sample. 
Typically, the channel impulse response (CIR), $\mathbf{h}$ is a random vector whose joint distribution is either empirically known or provided by standards bodies, e.g., the Extended Pedestrian-A (EPA) channel, the Tapped Delay Line (TDL) and the Clustered Delay Line (CDL) models~\cite{3GPP_CDL_channel}, etc.
In this paper, we assume $h_{\ell} \sim \mathcal{CN}(\mu_{\ell}, \sigma_{\ell}^2)$ in the analysis for Lemma~\ref{lemma:covariance_matrix_rank} and for part of the simulation results.
These statistics constitute the ``domain knowledge'' that is available during a basic ESN-based equalizer design.

\subsection{The Conventional Echo State Network (ESN)}
\label{sec:conventional_esn}
In this work, we consider the echo state network (ESN), a popular structure within the RC framework.
The ESN is typically defined as consisting of a ``reservoir'' containing $N_{\text{nodes}}$ randomly inter-connected neurons, along with an input weights layer and an output (readout) weights layer.
The parameters for this ESN structure are defined as follows:
\begin{itemize}
    \item $\mathbf{x}_{\text{in}}[n] \in \mathbb{C}^{d_{\text{in}}}$: ESN input at time index $n$.
    \item $\mathbf{x}_{\text{res}}[n] \in \mathbb{C}^{N_{\text{nodes}}}$: Reservoir state vector at time index $n$. 
    \item $\mathbf{W}_{\text{in}} \in \mathbb{C}^{N_{\text{nodes}} \times d_{\text{in}}}$: Input weights matrix.
    \item $\mathbf{W}_{\text{res}} \in \mathbb{C}^{N_{\text{nodes}} \times N_{\text{nodes}}}$: Reservoir weights matrix.
    \item $\mathbf{W}_{\text{out}} \in \mathbb{C}^{d_{\text{out}} \times N_{\text{nodes}}}$: Output weights matrix.
    \item $\mathbf{x}_{\text{out}}[n] \in \mathbb{C}^{d_{\text{out}}}$: ESN output at time index $n$. 
\end{itemize}
Here, $d_{\text{in}}$ and $d_{\text{out}}$ are the dimensions of the ESN's input and output respectively, typically interpreted as the number of features in the input and output. For a nonlinear activation function $\sigma(\cdot)$ (e.g., ReLU, tanh, etc.), the state update and output equations of the ESN are respectively given by:
\begin{align}
\mathbf{x}_{\text{res}}[n] &= \sigma \big(\mathbf{W}_{\text{res}}\mathbf{x}_{\text{res}}[n-1] + \mathbf{W}_{\text{in}}\mathbf{x}_{\text{in}}[n] \big), \\
\mathbf{x}_{\text{out}}[n] &= \mathbf{W}_{\text{out}}\mathbf{x}_{\text{res}}[n].
\label{output_eqn}
\end{align}
For analytical tractability, we consider a ``linear'' ESN so that $\sigma(\cdot)$ is an identity mapping. 
Such linearization allows analyzing the effectiveness of RC structures in a tractable manner, such as that in~\cite{Bollt2021}.
In conventional practice, only $\mathbf{W}_{\text{out}}$ is trained using a least squares approach, whereas $\mathbf{W}_{\text{in}}$ and $\mathbf{W}_{\text{res}}$ are randomly initialized from a particular pre-determined distribution and then kept fixed throughout the training and the test (inference) stages.
We hypothesize that the benefit of nonlinear activation in the reservoir and the optimization of $\mathbf{W}_{\text{in}}$ and $\mathbf{W}_{\text{res}}$ are either orthogonal or at least separable issues, as evidenced in our recent work~\cite{Jere2023WCL}. 

\section{Model Interpretability in RC}
\subsection{Signal Processing Preliminaries for the ESN}
\begin{figure}[h]
    \centering    \includegraphics[width=0.775\linewidth]{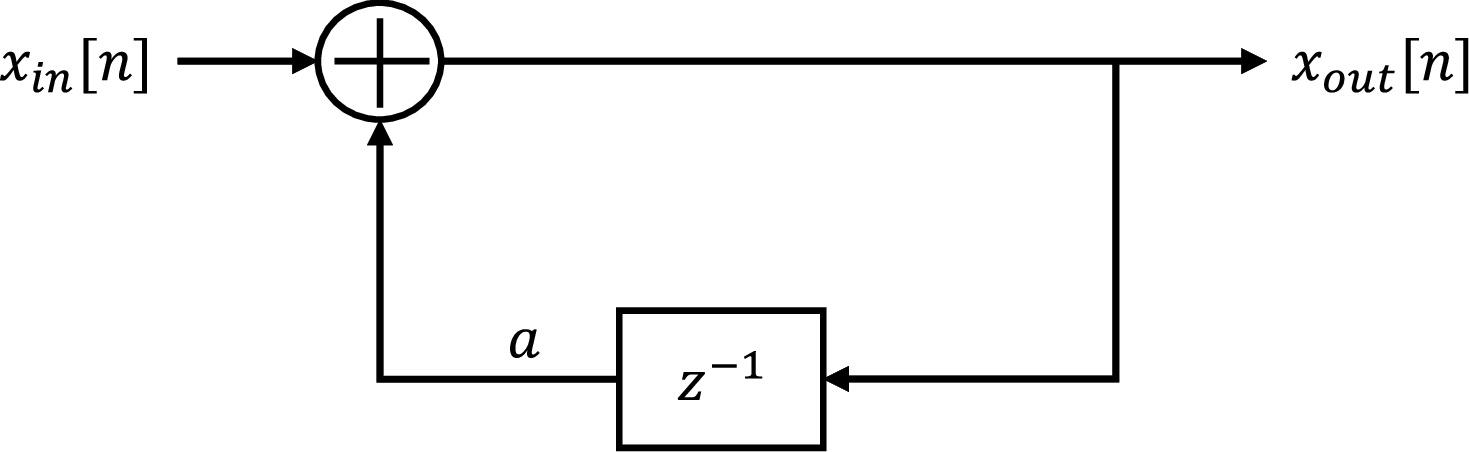}
    \caption{Modeling a neuron in the reservoir as a single-pole IIR filter.}
    \label{fig:single_pole_iir}
\end{figure}

Based on our previous work in~\cite{Jere2023WCL}, we know that an ESN with a single neuron and linear activation can be modeled as a first-order (single-pole) infinite impulse response (IIR) filter.
The block diagram showing the input-output relationship for a single neuron with a unit delay feedback loop is shown in Fig.~\ref{fig:single_pole_iir}.
For this filter, the system response is given by $H_{0}(z) = \sfrac{X_{\text{out}}(z)}{X_{\text{in}}(z)} = \frac{1}{1-az^{-1}}$.
In general, we can model the ESN's reservoir with $N$ neurons as an IIR filter with its system function given by $H_{\text{res}}(z) = \frac{\sum_{k=0}^M b_k z^{-k}}{1 + \sum_{k=1}^N a_k z^{-k}}$, where $M \leq N$.
This is an important foundation that will be used in Sec.~\ref{sec:optimum_esn_weights} to derive the optimum input and reservoir weights of the ESN given the statistical distribution of the channel impulse response coefficients.
For a given realization of the system response of the channel $H_{\text{ch}}(z)$, the objective of the equalizer is to use the available training data to ``learn'' the realization of the inverse response of the channel $\widehat{H}_{\text{inv}}(z)$ while minimizing the residual error, so that $\widehat{H}_{\text{inv}}(z) H_{\text{ch}}(z) \approx 1$. If the channel is minimum phase, then its direct inverse $\widehat{H}_{\text{inv}}(z) = 1/H_{\text{ch}}(z)$ is stable and the ESN-based equalizer can attempt to learn the corresponding frequency-domain inverse $1/H_{\text{ch}}(e^{j\omega})$. If the channel is non-minimum phase or mixed-phase, a stable inverse realization with the output incurring a delay must first be found, followed by the ESN-based equalizer learning this stable inverse while accounting for the delay in its output. In this paper, we investigate minimum phase channels, while the extension of this analysis and model optimization to non-minimum and mixed phase channels is part of future work.



\subsection{Optimum Orthogonal Basis Set}
\label{sec:optimum_basis_pca}
To characterize the channel in the frequency domain, we sample the digital frequency $\omega \in [0, 2\pi]$ rad/sample at $N$ unique points. Define a vector $\mathbf{v} \in \mathbb{C}^{N \times 1}$ whose $i$-th element is defined as $[\mathbf{v}]_i = H^{-1}_{\text{ch}}(j\omega_i) = 1/H_{\text{ch}}(j\omega_i)$, where $\omega_{i} = (\sfrac{2\pi}{N})i$ for $i=0,1,\ldots,N-1$. As the first step towards optimizing the traditionally untrained weights of the ESN-based detector, we will attempt to find a projection matrix $\mathbf{F} \in \mathbb{C}^{N \times M}$, which consists of $M$ orthonormal vectors $\{\mathbf{f}_m \}_{m=1}^M$ as its columns.
These $M$ vectors constitute a basis spanning a linear subspace $\mathcal{S}$, i.e. $\mathcal{S}= \myspan \left(\{\mathbf{f}_1, \mathbf{f}_2, \ldots, \mathbf{f}_M \} \right)$. 
Then, the low-dimensional representation of $\mathbf{v}$ is given by $\mathbf{u} = \mathbf{F}^H \mathbf{v}$, where $\mathbf{u} \in \mathbb{C}^M$ is known as the \emph{latent vector}~\cite{Murphy2012}.
Subsequently, the reconstruction of the channel inverse $\widehat{\mathbf{v}} \in \mathbb{C}^{N \times 1}$ using the low-dimensional projection $\mathbf{u}$ is given by $\widehat{\mathbf{v}} = \mathbf{F} \mathbf{u} = \mathbf{F} \mathbf{F}^H \mathbf{v}$.
Note that while $\mathbf{F}^H \mathbf{F} = \mathbf{I}_M$, $\mathbf{F} \mathbf{F}^H \neq \mathbf{I}_N$. Effective equalization of the distortion caused by the wireless channel translates to minimizing the projection error $\varepsilon_{\mathrm{proj}}$, where $\varepsilon_{\mathrm{proj}} = \mathbb{E}_{\mathbf{v}}[\| \mathbf{v} - \widehat{\mathbf{v}} \|_{2}^{2}]$.
It can be shown~\cite{Murphy2012} that this projection error 
is minimized by setting $\mathbf{F} = \mathbf{Q}_M$, where $\mathbf{Q}_M$ contains the $M$ eigenvectors of the covariance matrix $\mathbb{E}[(\mathbf{v} - \mathbb{E}[\mathbf{v}]) (\mathbf{v} - \mathbb{E}[\mathbf{v}])^H]$ in decreasing order of the corresponding eigenvalues.
This result follows from
the widely used dimensionality reduction technique of principal component analysis (PCA), which is a special case of the more general Karhunen-Lo\'eve (KL) expansion for stochastic processes.
Since we assume knowledge of the channel statistics,
we will use a fixed but sufficiently large number of realizations drawn from the known joint probability distribution $p_{\mathbf{h}}$ of the channel tap coefficients. 
Denoting $N_{\text{obs}}$ as the number of realizations observed of the channel, the empirical covariance matrix $\widehat{\mathbf{\Sigma}} \in \mathbb{C}^{N \times N}$ can be computed as $\widehat{\mathbf{\Sigma}} = \frac{1}{N_{\text{obs}}} \sum_{c=1}^{N_{\text{obs}}} \big(\mathbf{v}_c - \mathbb{E}[\mathbf{v}_c]\big) \big(\mathbf{v}_c - \mathbb{E}[\mathbf{v}_c] \big)^H$,
where $\mathbf{v}_c$ is the $c$-th realization of the frequency domain channel inverse.
Therefore, the empirical approximation of the optimum basis vectors can be obtained by the eigendecomposition of $\widehat{\mathbf{\Sigma}}$, i.e., $\widehat{\mathbf{\Sigma}} = \mathbf{Q} \mathbf{D} \mathbf{Q}^{-1}$,
where $\mathbf{Q} \in \mathbb{C}^{N \times N}$ is an orthogonal matrix containing the eigenvectors of $\widehat{\mathbf{\Sigma}}$ in its columns and $\mathbf{D} = \diag(\lambda_1, \lambda_2, \ldots, \lambda_N) \in \mathbb{C}^{N \times N}$ is the diagonal matrix of the corresponding eigenvalues.
Finally, the projection matrix $\mathbf{F}$ is simply given by $\mathbf{F} = \mathbf{Q}_M \in \mathbb{C}^{N \times M}$, where $\mathbf{Q}_M$ contains the $M$ eigenvectors in the decreasing order of eigenvalues. Since $\mathbf{v}$ is complex-valued, obtaining the corresponding optimum complex-valued basis vectors in $\mathbf{F}$ requires special handling, which is detailed in Appendix~\ref{sec:Appendix_A}.
Therefore, it is clear that the choice of $M$ plays an important role in lowering the reconstruction error.
In the following lemma, we present an analytical characterization of the dimensionality of the subspace spanned by the frequency-domain channel inverse $\mathbf{v}$ that the ESN attempts to approximate. Having defined $\widetilde{\mathbf{v}} \in \mathbb{R}^{2N} = [\Re(\mathbf{v}) \; \Im(\mathbf{v})]^T$, the lemma provides guidance on the number of significant eigenvectors needed in $\mathbf{F}$ for an approximation error on the order of $\mathcal{O}\left(\sigma_{0}^{4} \right)$.
\begin{lemma}
    For an $L$-tap channel impulse response $\mathbf{h}$ following an i.i.d. Gaussian distribution, the covariance matrix of $\Tilde{\mathbf{v}}$ has an $\epsilon$-rank given by $\rank_{\epsilon} \left( \mathbb{E}[(\Tilde{\mathbf{v}} - \mathbb{E}[\Tilde{\mathbf{v}}]) (\Tilde{\mathbf{v}} - \mathbb{E}[\Tilde{\mathbf{v}}])^T] \right) = 4(L+1)$, where $\epsilon = \mathcal{O}(\sigma_0^4)$. 
    \label{lemma:covariance_matrix_rank}
\end{lemma}
The complete proof of this result is provided in Appendix~\ref{sec:Appendix_B}. 
We remark that the number of significant eigenvectors $M$ required to achieve a $\mathcal{O}(\sigma_{0}^{4})$ reconstruction MSE is at least $4(L+1)$. 
Thus, we need $M \sim \mathcal{O}(L)$ to achieve a reconstruction error comparable to $\sigma_0^4$.
This important conclusion not only provides a guide to heuristically choosing a suitable dimensionality of the optimum basis set, but also influences the reservoir size in neurons, as discussed in Sec.~\ref{sec:optimum_esn_weights} next.

\subsection{From Optimum Basis to Optimum ESN Design}
\label{sec:optimum_esn_weights}
The KL basis expansion method, specifically the PCA dimensionality reduction procedure of Sec.~\ref{sec:optimum_basis_pca} provides the mean square error (MSE) minimizing optimum basis set in the time-domain equalization of the wireless channel. To connect this with the optimum design of the ESN, we revert to the linearized ESN model. 
We will experimentally show in Sec.~\ref{sec:performance_evaluation} that the optimum weights derived from PCA also in fact, apply to the conventional ESN with nonlinear activation.
Using the technique described in Sec. IV-D, we arrive at $\mathbf{F} = [\mathbf{f}_1 \; \mathbf{f}_2 \; \ldots \; \mathbf{f}_M]$, where $\mathbf{f}_{m} \in \mathbb{C}^{M \times 1}$ is the $m^{\text{th}}$ column in the projection matrix $\mathbf{F}$. The next step is key as it relates the orthogonal eigenvectors $\mathbf{f}_m$ to the process of ESN construction. Denoting the continuous frequency version of $\mathbf{f}_m$ as $f_m(j\omega)$, we approximate $f_m(j\omega)$ for $m=1,\ldots,M$ as a proper rational polynomial (RP) given by
\begin{align}
    f_m(j\omega) \approx \frac{\sum_{k=0}^{K'} c_{m,k} e^{-j\omega k}}{1 + \sum_{k=1}^{K} d_{m,k} e^{-j\omega K}} \overset{\mtrian}{=} R_{m}(j\omega),
\end{align}
with the constraint $K' < K$. 
This approximation is motivated from the claim in Sec. IV-A that an ESN can be modeled as a general IIR filter with its transfer function given by a rational polynomial.
The coefficients $\{c_{m,k}\}_{k=0}^{K'}$ and $\{d_{m,k}\}_{k=1}^{K}$ can be found by solving the linear system given by $\mathbf{f}_{m} = \big[\mathbf{\Omega}_1 \;\; \mathbf{\Omega}_2 \big] \mathbf{c}_m$,
where $\mathbf{\Omega}_1 \in \mathbb{C}^{N \times (K'+1)}$ and $\mathbf{\Omega}_2 \in \mathbb{C}^{N \times K}$ are defined row-wise as $\mathbf{\Omega}_{1}^{(r)} \in \mathbb{C}^{1 \times (K'+1)} \overset{\mtrian}{=} [1 \;\; e^{-j\omega_r} \;\; \ldots \;\; e^{-jK'\omega_r}]$ and $\mathbf{\Omega}_{2}^{(r)} \in \mathbb{C}^{1 \times K} \overset{\mtrian}{=} [-f_m(j\omega_r)e^{-j\omega_r} \;\; \ldots \;\; -f_m(j\omega_r)e^{-jK\omega_r}]$,
and $\mathbf{c}_m \overset{\mtrian}{=} [c_{m,0} \; c_{m,1} \; \ldots \; c_{m,K'} \; d_{m,1} \; \ldots \; d_{m,K}]^T \in \mathbb{C}^{K+K'+1}$. 
Therefore, the coefficients of the numerator and the denominator polynomials can be found as $\mathbf{c}_m = \big[\mathbf{\Omega}_1 \; \; \mathbf{\Omega}_2 \big]^{\dagger} \mathbf{f}_m$,
where $(\cdot)^{\dagger}$ denotes the Moore-Penrose matrix pseudoinverse. 

Next, we decompose each rational polynomial $R_m(j\omega)$ into a sum of first-order IIR transfer function forms according to 
\begin{align}
    R_m(j\omega) \approx \sum_{k=1}^{K} \frac{q_{m,k}}{1-p_{m,k}e^{-j\omega}}, \; m=1,2,\ldots,M. 
    \label{eq:rp_decomposition}
\end{align}
From Eq.~\eqref{eq:rp_decomposition}, we can recognize that each term in the sum on the RHS is the frequency response of a single-pole IIR filter, i.e., a single neuron with a unit delay feedback loop. Thus, the rational polynomial approximation $R_m(j\omega)$ of the $m^{\text{th}}$ ``principal component'' can be decomposed as a sum of the frequency responses of $K$ independent neurons in the reservoir. This translates to a collection of $K$ non-interconnected neurons, each with a unit delay feedback loop. For a total of $M$ principal components (eigenvectors), this amounts to having a total of $N_{\text{nodes}} = MK$ non-interconnected neurons in the reservoir.
The choice of the polynomial order $K$ is important, in that it must be heuristically chosen so that the approximation error $\int_{0}^{2\pi} |f_{m}(j\omega) - R_{m}(j\omega)|^2 d\omega$ is acceptable.

\begin{figure*}[htbp]
    \centering    \includegraphics[width=0.7268\linewidth]{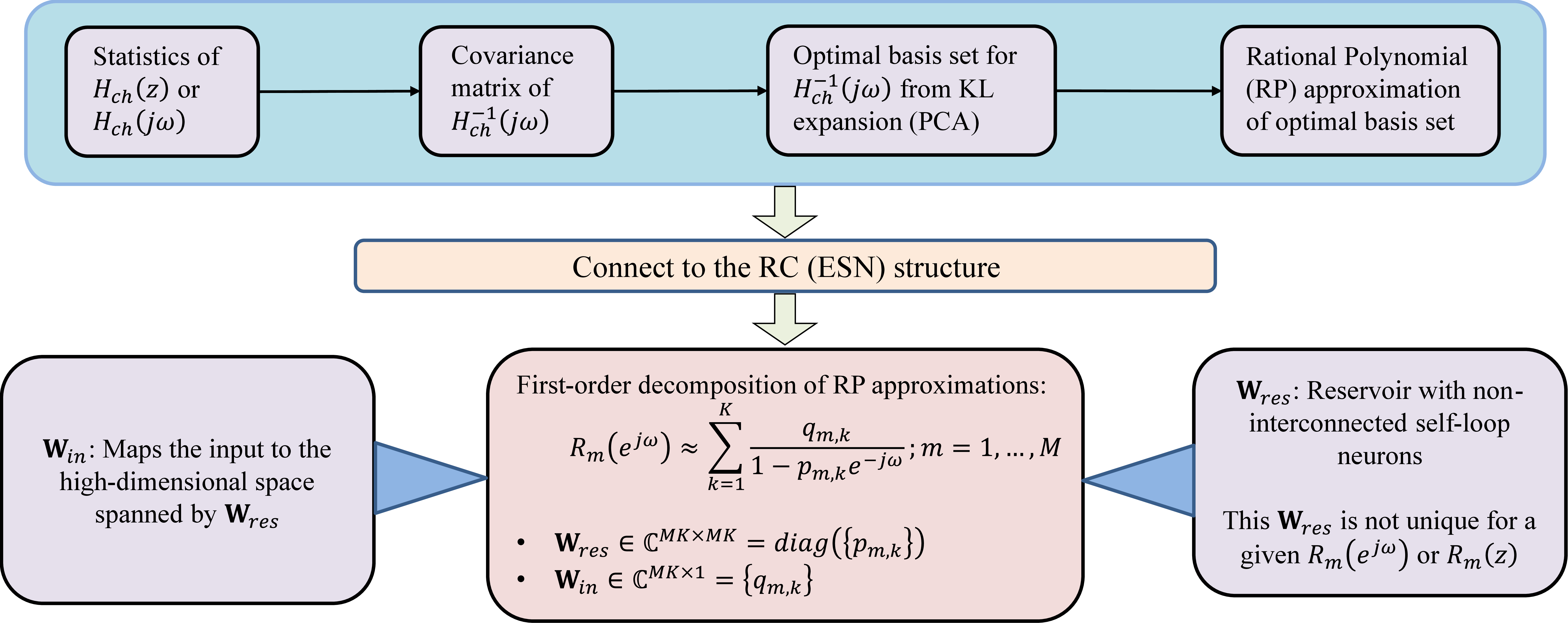}
    \caption{Flowchart outlining the derivation of ``optimum'' ESN model weights (conventionally untrained) from available channel statistics.}
    \label{fig:Flowchart}
\end{figure*}

Finally, we relate this analysis to the model weights of the ESN. Specifically, the input weights vector and the reservoir weights matrix are respectively given by
\begin{align}
    \mathbf{W}_{\text{in}} = \myvec(\{q_{m,k} \}) \in \mathbb{C}^{MK \times 1} \label{eq:opt_Win}, \\
    \mathbf{W}_{\text{res}} = \diag(\{p_{m,k} \}) \in \mathbb{C}^{MK \times MK} \label{eq:opt_Wres}.
\end{align}
Therefore, starting with the available domain knowledge in the form of channel statistics, we have arrived at the ``optimum'' ESN input weights and reservoir weights, as given by Eq.~\eqref{eq:opt_Win} and Eq.~\eqref{eq:opt_Wres} respectively. 
The complete procedure of deriving the ``optimum'' untrained weights of the ESN provided channel statistics are available, is depicted in the flowchart of Fig.~\ref{fig:Flowchart}.
This provides a path toward incorporating domain knowledge directly into the design of the ESN instead of initializing the untrained weights $\mathbf{W}_{\text{in}}$ and $\mathbf{W}_{\text{res}}$ from a uniform or Gaussian distribution, as per conventional practice.

\section{Experimental Validation}
\label{sec:performance_evaluation}
In this section, we present simulation results validating the optimal design of the ESN-based symbol detector. We use the Symbol Error Rate (SER) as the metric for evaluating the receive processing performance with input and reservoir weights randomly drawn from a uniform distribution compared with the optimum weights derived using the knowledge of the channel statistics. We consider two sets of channel distributions: i) A channel impulse response (CIR) with an exponentially decaying power delay profile (PDP) and ii) 3GPP's Clustered Delay Line (CDL) models~\cite{3GPP_CDL_channel} `CDL-D' and `CDL-E'.
For (i), we model the complex-valued channel tap coefficients as $\mathbf{h} = [1 \; h_1 \; h_2 \; \ldots \; h_{L-1}]$, where $h_{\ell}$ is modeled as a Gaussian random variable with the distribution $h_\ell \sim \mathcal{CN}(\mu_{\ell}, \sigma_{\ell}^2)$ where $\mu_{\ell} = e^{-1.5\ell} \mu_0, \sigma_{\ell}^{2} = \frac{1}{2}e^{-3.75\ell}$  
for $\ell = 1, \ldots, L-1$, with $\mu_0 \sim \mathcal{CN}(0,1)$ being a circularly symmetric Gaussian random variable. 
The realizations for the CDL-D and CDL-E models are generated as per their parameters specified in~\cite{3GPP_CDL_channel}.
The ESN implemented in these simulations is the basic ESN without the structural improvements made in our previous works, e.g., windowing, concatenated skip connection, delay learning~\cite{zhou2020_LTD}, knowledge of the modulation constellation~\cite{RCstruct}, etc.
In addition, we employ the hyperbolic tangent (tanh) nonlinear activation following conventional deep learning practice, and demonstrate that our analysis of the linear scenario transfers over gracefully to the scenario with nonlinear activation as well.



\subsection{Improvement in Symbol Error Rate (SER) Performance}
In order to verify the optimality of the ESN model weights derived using the procedure of Sec.~\ref{sec:optimum_esn_weights}, we simulate a point-to-point simulation of a transceiver in a SISO-OFDM system.
For the exponentially decaying PDP channel, we use $L=10$ taps, and use $M=L=10$ eigenvectors in line with the guidance of Lemma~\ref{lemma:covariance_matrix_rank}, with $K=10$ for the rational polynomial approximation of each polynomial resulting in $N_{\text{nodes}}=MK=100$ neurons in the reservoir. For the CDL-D channel which has $L=14$ taps (modeled from clusters), we also use $M=L=14$ and $K=10$ giving $N_{\text{nodes}}=140$. Finally for the CDL-E channel with $L=15$ taps, we also use $M=L=15$ and $K=10$ resulting in $N_{\text{nodes}}=150$. 
We use $K'=K-1$ in all RP approximations.
For the ESN models with randomly initialized $\mathbf{W}_{\text{in}}$ and $\mathbf{W}_{\text{res}}$, the spectral radius of the ESN is set to $0.4$ and the sparsity to $0.6$.
The OFDM parameters are set as FFT size = $1024$, cyclic prefix (length) = $160$, number of pilot OFDM symbols = $4$ and number of data (payload) OFDM symbols = $13$ in each subframe. 

\begin{figure}[htbp]
    \centering    \includegraphics[width=\linewidth]{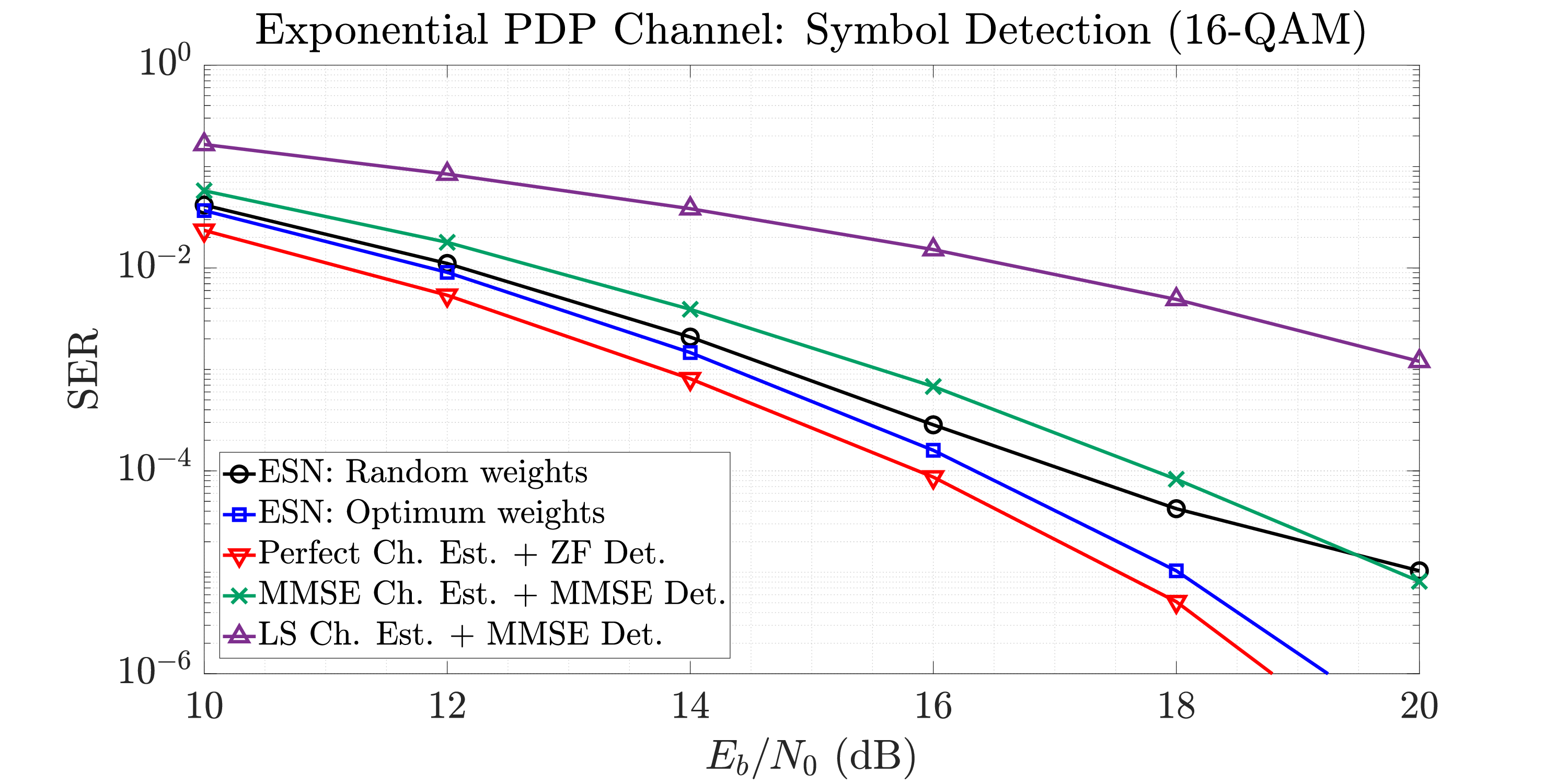}
    \caption{Symbol detection performance comparison for exponentially decaying PDP channel with 16-QAM modulation.}
    \label{fig:SER_custom_16qam_nonlinear}
\end{figure}

\begin{figure}[htbp]
    \centering    \includegraphics[width=\linewidth]{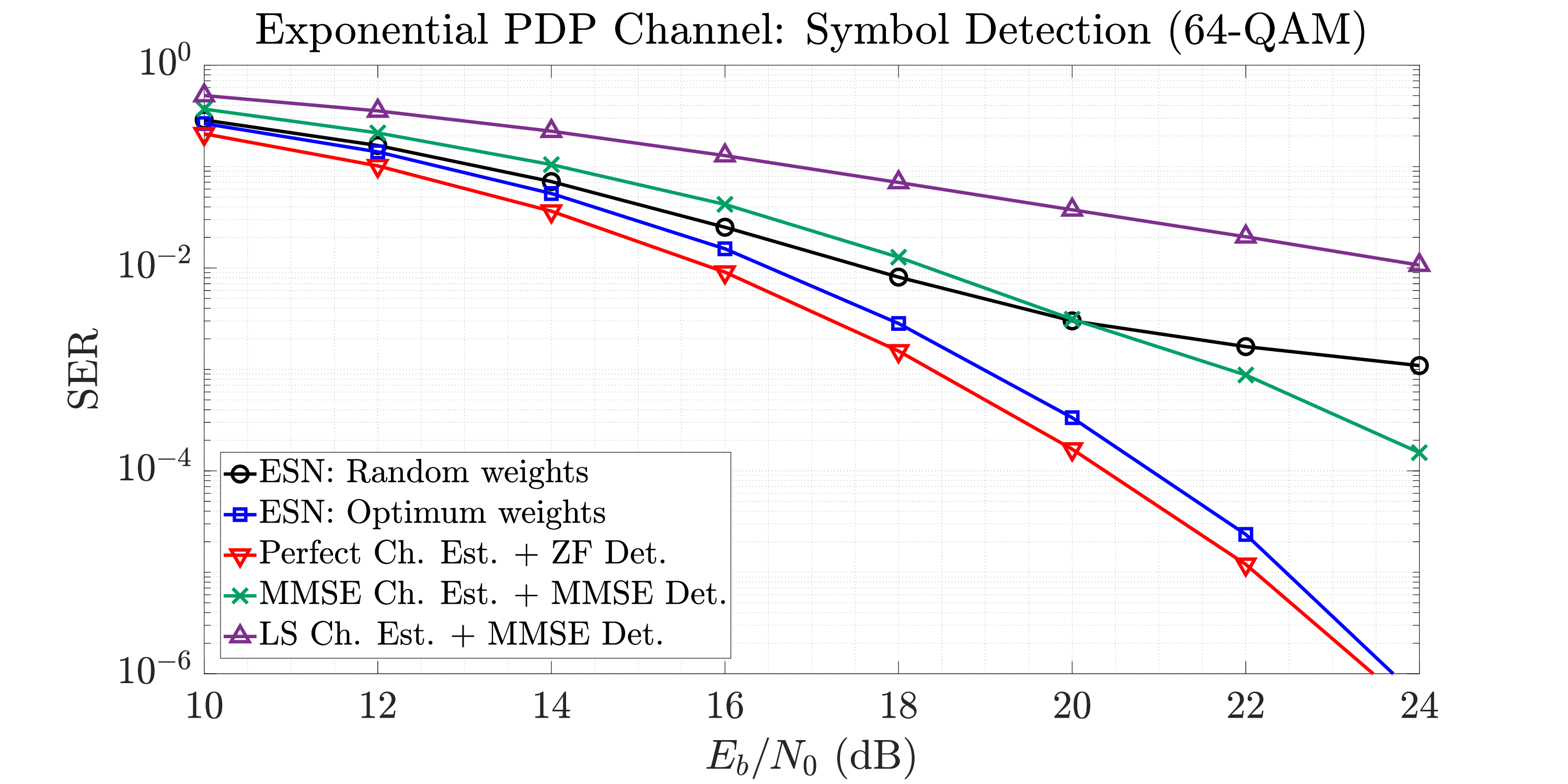}
    \caption{Symbol detection performance comparison for exponentially decaying PDP channel with 64-QAM modulation.}
    \label{fig:SER_custom_64qam_nonlinear}
\end{figure}

\begin{figure}[htbp]
    \centering    \includegraphics[width=\linewidth]{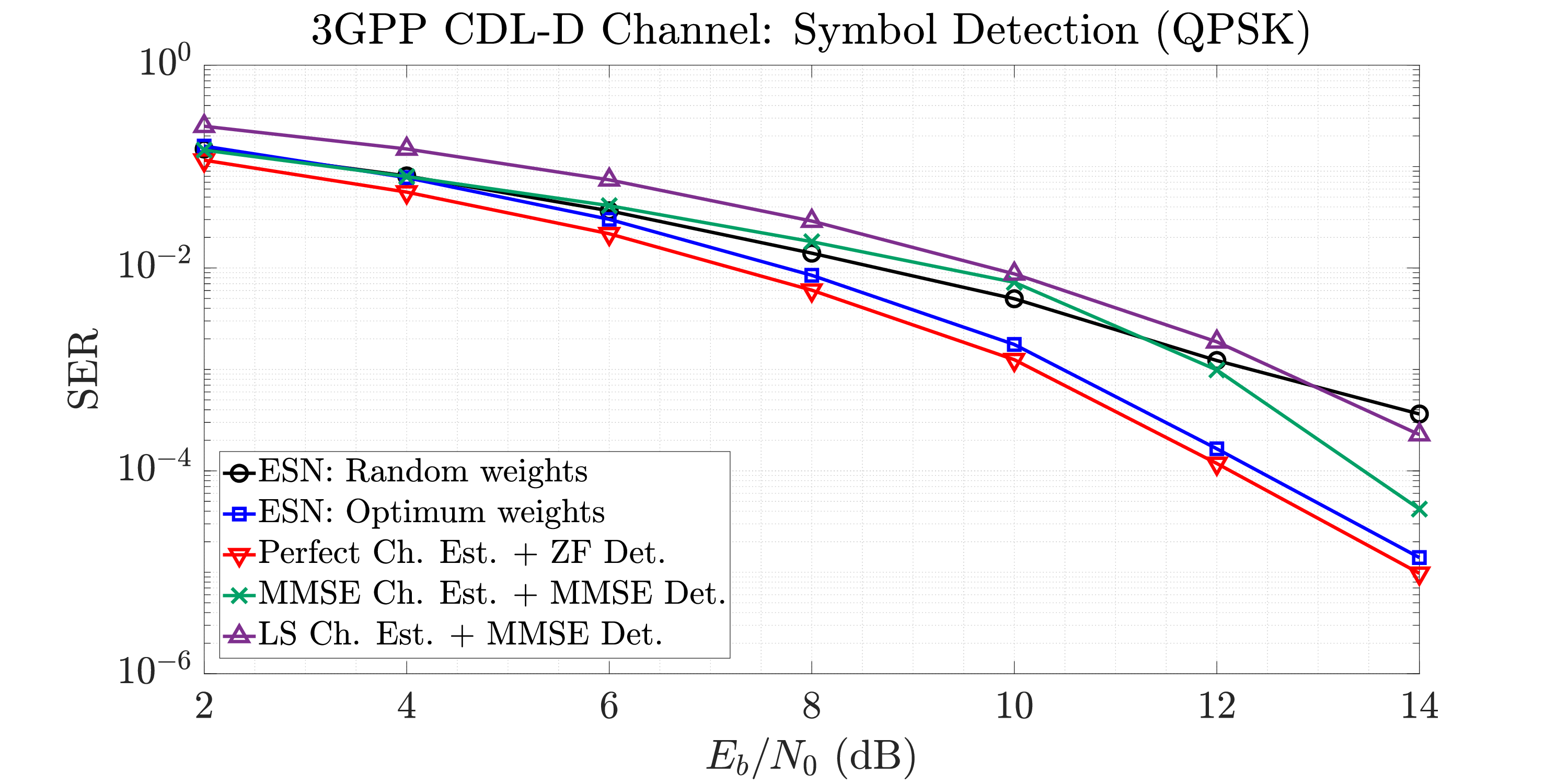}
    \caption{Symbol detection performance comparison for 3GPP CDL-D channel with QPSK modulation.}
    \label{fig:SER_CDL-D_nonlinear}
\end{figure}

\begin{figure}[htbp]
    \centering    \includegraphics[width=\linewidth]{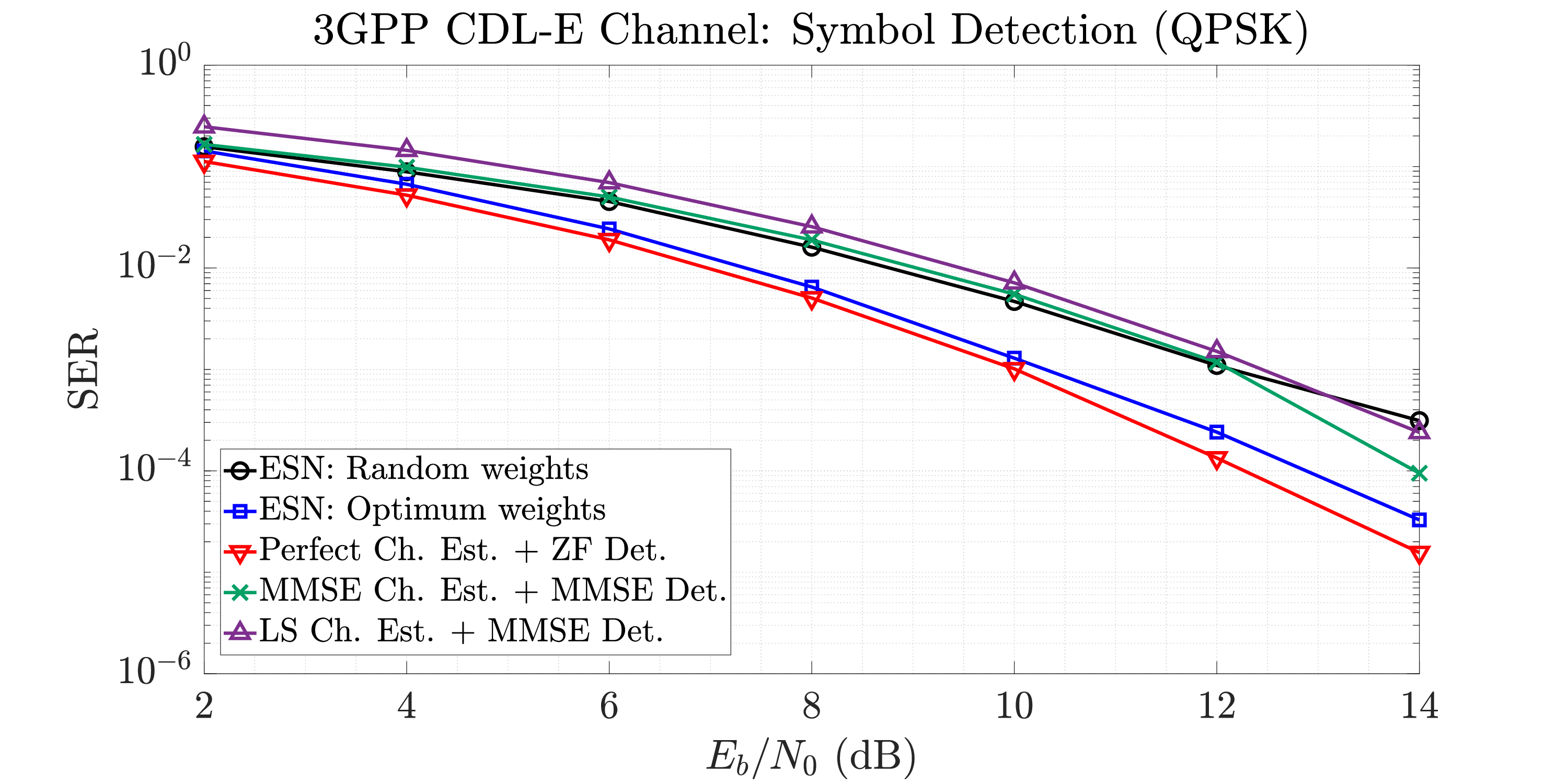}
    \caption{Symbol detection performance comparison for 3GPP CDL-E channel with QPSK modulation.}
    \label{fig:SER_CDL-E_nonlinear}
\end{figure}

In these simulations, $\mathbf{W}_{\text{out}}$ is trained via pseudoinverse-based least squares using the limited training data in the form of pilot OFDM symbols (following a block pilot structure), consistent with our previous work~\cite{zhou2020_LTD,zhou2020rcnet, RCstruct}.
The non-learning baseline approaches include: i) perfect channel knowledge at the receiver with zero-forcing (ZF) equalization per-subcarrier, ii) empirical minimum mean square error (MMSE) channel estimation with MMSE equalization, and iii) least squares (LS) channel estimation with MMSE equalization.
We can observe from Fig.~\ref{fig:SER_custom_16qam_nonlinear} that when $\mathbf{W}_{\text{in}}$ and $\mathbf{W}_{\text{res}}$ are chosen according to Eq.~\eqref{eq:opt_Win} and Eq.~\eqref{eq:opt_Wres} respectively, the SER performance with 16-QAM modulation is significantly better compared to the case when $\mathbf{W}_{\text{in}}$, $\mathbf{W}_{\text{res}}$ are randomly drawn from $(\mathcal{U}(-1, 1) + j\mathcal{U}(-1, 1))$. The improvement in performance is even more significant for higher-order modulation such as 64-QAM as seen in Fig.~\ref{fig:SER_custom_64qam_nonlinear}, where the model with random weights exhibits an error floor at higher $E_b/N_0$.
In addition, the SER curves with QPSK modulation under 3GPP standards-compliant channels such as CDL-D and CDL-E are also shown in Fig.~\ref{fig:SER_CDL-D_nonlinear} and Fig.~\ref{fig:SER_CDL-E_nonlinear} respectively.
The error floor behavior is also clearly visible in the case of random weights, whereas the model with optimum weights can achieve SER close to the ideal curve with perfect channel knowledge.
This paves the way for enhanced reliability of NN-based receivers in contested and interference-limited environments.



\section{Conclusion and Future Work}
In this work, we present a first principles-based method to understand a widely used category of reservoir computing, the echo state network. 
Building on prior work, a geometric view of the channel equalization task using the ESN as a projection problem is developed. 
Most importantly, a conceptual bridge is established between available domain knowledge in the form of channel statistics and the optimum initialization of the conventionally untrained weights of the ESN. 
The validity of this optimized initialization is experimentally demonstrated via simulations of receive symbol detection performance, paving the path for model interpretability of RC in wireless applications and is a significant first step towards explainable machine learning and improved reliability with NN techniques in challenging wireless environments. 
For future work, this analysis will be extended to non-minimum phase and MIMO channels to develop a fully analytical characterization of the optimum distribution of the ESN model weights, going beyond the heuristic development in this work.

\begin{appendices}

\section{Construction of Complex-Valued Optimum Basis}
\label{sec:Appendix_A}
First, we stack the real and imaginary parts of $\mathbf{v}$ vertically in $\widetilde{\mathbf{v}} \in \mathbb{R}^{2N}$.
The empirical covariance matrix $\widetilde{\mathbf{\Sigma}} \in \mathbb{R}^{2N \times 2N}$ for $\widetilde{\mathbf{v}}$ is computed similar to that for $\mathbf{v}$.
Next, the eigendecomposition for $\widetilde{\mathbf{\Sigma}}$ is performed according to $\widetilde{\mathbf{\Sigma}} = \widetilde{\mathbf{Q}} \widetilde{\mathbf{D}} \widetilde{\mathbf{Q}}^{-1}$,
where $\widetilde{\mathbf{Q}}_M \in \mathbb{R}^{2N \times M}$ contains in its columns, the eigenvectors corresponding to the $M$ largest eigenvalues, and $\widetilde{\mathbf{D}} = \diag(\widetilde{\lambda}_1, \ldots, \widetilde{\lambda}_{2N})$ contains the corresponding eigenvalues with $\{\widetilde{\lambda}_i\}_{i=1}^{2N} \in \mathbb{R}$. Finally, the optimum complex-valued basis vectors in $\mathbf{F} \in \mathbb{C}^{N \times M}$ are given by $\mathbf{F} = \mathbf{Q}_M = \mathbf{P} \widetilde{\mathbf{Q}}_M$,
where $\mathbf{P} \in \mathbb{C}^{N \times 2N}$ is defined as $\mathbf{P} = \begin{bmatrix}
\mathbf{1}_N & j\mathbf{1}_N
\end{bmatrix}$.

\section{Proof of Lemma~\ref{lemma:covariance_matrix_rank}}
\label{sec:Appendix_B}
Consider the frequency response of an $L$-tap fading channel with random i.i.d. tap weights given by $H_{\text{ch}}(e^{j\omega})=\sum_{\ell=0}^{L-1}(b_{\ell}+\mu_{\ell})e^{-j\omega \ell}$,
where $b_{\ell} \sim \mathcal{CN}  (0,\sigma_{\ell}^2)$ and $\mu_{\ell} \in \mathbb{C}$ is the constant mean of the $\ell^{\text{th}}$ tap.
Discretizing $H_{\text{ch}}(j\omega)$ at $N$ points in frequency, we get    $\omega_n = 2\pi n/N$, where $n=0,\ldots,N-1$.
With slight abuse of notation, we denote the $n^{\text{th}}$ element of the sampled $[H_{\text{ch}}(e^{j\omega})] \in \mathbb{C}^N$ as $[H_{\text{ch}}(e^{j\omega})]_{\omega=\omega_n} = H_{\text{ch}}(e^{j\omega_n}) \overset{\mtrian}{=}  H_n=\sum_{\ell=0}^{L-1}(b_{\ell}+\mu_{\ell})e^{-j2\pi n\ell /N}$.
Let $\mathbf{b} \in \mathbb{C}^L \overset{\mtrian}{=}[b_0,..,b_{L-1}]^T$ and $\boldsymbol \mu \in \mathbb{C}^L \overset{\mtrian}{=} [\mu_0,..,\mu_{L-1}]^T$, we define the channel inverse vector in the frequency domain as $\mathbf{v}=[\sfrac{1}{H_0},..,\sfrac{1}{H_{N-1}} ]^T$. 
In the following analysis, we will use the multi-variable Taylor expansion which requires a function to be analytic.
However, $\mathbf{v}$ is a vector of complex-valued functions and its covariance is not an analytic function. Therefore, we find a purely real-valued equivalent of the covariance matrix.
Define the real equivalent vector 
$\widetilde{\mathbf{v}} \in \mathbb{R}^{2N} = [\Re(\mathbf{v}) \; \Im(\mathbf{v})]^T$. 
To find the rank of the covariance matrix $\mathbf{K}_{\Tilde{v}\Tilde{v}} \overset{\mtrian}{=}  \mathbb{E}[(\widetilde{\mathbf{v}} - \mathbb{E}[\widetilde{\mathbf{v}}]) (\widetilde{\mathbf{v}} - \mathbb{E}[\widetilde{\mathbf{v}}])^T]$, it suffices to find the rank of the correlation matrix $\mathbf{R}_{\Tilde{v}\Tilde{v}} \overset{\mtrian}{=} \mathbb{E}[\widetilde{\mathbf{v}}\widetilde{\mathbf{v}}^{T}]$. 
First, we analyze
the structure of the matrix 
$\widetilde{\mathbf{v}}\widetilde{\mathbf{v}}^{T} 
\in \mathbb{R}^{2N \times 2N}
$ as
\begin{align}\label{re_im_stack_gramm} \widetilde{\mathbf{v}}\widetilde{\mathbf{v}}^{T} 
= \begin{bmatrix}
 \Re \left(\mathbf{v}\mathbf{v}^{T}\right) & \hspace{-2mm} \Im\left( \mathbf{v}\mathbf{v}^{T}\right) \\
\Im \left( \mathbf{v}\mathbf{v}^{T}\right) & \hspace{-2mm} -\Re \left(\mathbf{v}\mathbf{v}^{T}\right)\\
\end{bmatrix} 
+\begin{bmatrix}
 \Re \left(\mathbf{v}\mathbf{v}^{H}\right) & \hspace{-2mm} -\Im\left( \mathbf{v}\mathbf{v}^{H}\right) \\
-\Im \left( \mathbf{v}\mathbf{v}^{H}\right) & \hspace{-2mm} \Re \left(\mathbf{v}\mathbf{v}^{H}\right)\\
\end{bmatrix}.
\end{align}
From \eqref{re_im_stack_gramm}, the correlation matrix $\mathbf{R}_{\Tilde{v}\Tilde{v}}$ can be expressed as
\begin{align}
\mathbf{R}_{\Tilde{v}\Tilde{v}} 
= 2\begin{bmatrix}
 \mathbb{E} \left[ \Re(\mathbf{v})\Re(\mathbf{v}^{T})\right] & j\mathbb{E}\left[ \Im(\mathbf{v})\Im(\mathbf{v}^{T})\right] \\
j\mathbb{E}\left[ \Im(\mathbf{v})\Im(\mathbf{v}^{T})\right] &  -j\mathbb{E} \left[ \Re(\mathbf{v})\Im(\mathbf{v}^{T})\right]  \\
\end{bmatrix}.
\end{align}
Now, we pursue a bound on the rank of $\mathbf{R}_{\Tilde{v}\Tilde{v}}$ by first finding a bound on the rank of each of its  submatrices $\mathbb{E} \left[ \Re(\mathbf{v})\Re(\mathbf{v}^{T})\right]$, $\mathbb{E}\left[ \Im(\mathbf{v})\Im(\mathbf{v}^{T})\right]$ and $\mathbb{E} \left[ \Re(\mathbf{v})\Im(\mathbf{v}^{T})\right]$.
The $(n,m)$-th element of $\left[ \Re(\mathbf{v})\Re(\mathbf{v}^{T})\right] \in \mathbb{R}^{N \times N}$, denoted as  $\left[ \Re(\mathbf{v})\Re(\mathbf{v}^{T})\right]_{n,m} = \left[\Re({H}_n^{-1})\Re({H}_m^{-1})\right]$ can be treated as a real-valued multivariate function $
f(\widetilde{\mathbf{b}}) \overset{\mtrian}{=} \left[\Re({H}_n^{-1})\Re({H}_m^{-1})\right] : \mathbb{R}^{2L} \rightarrow \mathbb{R}$,  where $\widetilde{\mathbf{b}}=[b_0^{(\mathrm{re})},..b_{L-1}^{(\mathrm{re})} b_0^{(\mathrm{im})},..b_{L-1}^{(\mathrm{im})}]^T \in \mathbb{R}^{2L}$  and $b_{\ell}^{(\mathrm{re})} = \Re(b_{\ell})$, $b_{\ell}^{(\mathrm{im})} = \Im(b_{\ell})$.
Applying the multivariate Taylor expansion up to the second power about $\widetilde{\mathbf{b}}=\mathbf{0}$, we get
\begin{align}\label{re_hn_inv_hm_inv}
f(\widetilde{\mathbf{b}})\approx f(\mathbf{0}) + \left[ \nabla f(\mathbf{0})\right]^T\widetilde{\mathbf{b}} +  0.5\widetilde{\mathbf{b}}^T \left[ \nabla^2 f(\mathbf{0}) \right] \widetilde{\mathbf{b}},
\end{align}
where $\left[\nabla f(\mathbf{0}) \right]_i = \frac{\partial}{\partial \tilde{b}_i}\left[\Re({H}_n^{-1})\Re({H}_m^{-1})\right](\mathbf{0})$ and  $\left[\nabla^2 f(\mathbf{0})\right]_{i,j} = \frac{\partial}{\partial \tilde{b}_i \partial \tilde{b}_j}\left[\Re({H}_n^{-1})\Re({H}_m^{-1})\right](\mathbf{0})$ are the Jacobian vector and the Hessian matrix respectively.
Taking the expectation of \eqref{re_hn_inv_hm_inv} and since $\mathbb{E}[\widetilde{b}_i]=0$, it follows that
\begin{align}
\label{E_re_hn_inv_hm_inv}
&\mathbb{E}\left[f(\widetilde{\mathbf{b}})\right]\approx \Re\left(\frac{1}{\boldsymbol{\mu}^T\boldsymbol{\lambda}_n  }\right)\Re\left(\frac{1} {\boldsymbol{\mu}^T\boldsymbol{\lambda}_m  }\right) \nonumber \\
&+ \sum_i \left[\nabla f(\mathbf{0})\right]_i\mathbb{E}[\widetilde{b}_i] +  \frac{1}{2} \sum_{i=0}^{2L-1}\sum_{j=0}^{2L-1} \left[\nabla^2 f(\mathbf{0})\right]_{i,j} \mathbb{E}[\widetilde{b}_i\widetilde{b}_j] \nonumber \\
&= \Re\left(\frac{1}{\boldsymbol{\mu}^T\boldsymbol{\lambda}_n  }\right)\Re\left(\frac{1}{\boldsymbol{\mu}^T\boldsymbol{\lambda}_m  }\right) \nonumber \\
&+ \frac{1}{2} \sum_{i=0}^{2L-1} \left[\frac{\partial^2}{\partial \tilde{b}_i^2}\left[\Re({H}_n^{-1})\Re({H}_m^{-1})\right](\mathbf{0})\right] \sigma_i^2,
\end{align}
where $[\boldsymbol{\lambda}_n]_{\ell} \overset{\mtrian}{=} e^\frac{-j2\pi n \ell}{N}$. 
Expanding $\frac{\partial^2}{\partial \widetilde{b}_i^2}\left[\Re({H}_n^{-1})\Re({H}_m^{-1})\right](\mathbf{0}) = [\nabla^{2}f(\mathbf{0})]_i$ and noting that the partial derivative and real part operators can be interchanged,
\begin{align}
\label{eq:double_derivative}
&[\nabla^{2}f(\mathbf{0})]_i \nonumber \\
&= \Bigg[\Re\left(\frac{\partial^2({H}_n^{-1})}{\partial \tilde{b}_i^2}\right) \Re({H}_m^{-1}) 
+\Re({H}_n^{-1})\Re \left(\frac{\partial^2 ({H}_m^{-1})}{\partial \tilde{b}_i^2} \right) \nonumber \\
&+2\Re \left(\frac{\partial ({H}_n^{-1})}{\partial \tilde{b}_i} \right) \Re \left(\frac{\partial ({H}_m^{-1})}{\partial \tilde{b}_i} \right)\Bigg](\mathbf{0}).
\end{align}
Thus, we get $\frac{\partial ({H}_n^{-1})}{\partial \widetilde{b}_i} = -\frac{1}{H_n^2} \frac{\partial {H}_n}{\partial \widetilde{b}_i}|_{\widetilde{\mathbf{b}}=\mathbf{0}} = -\frac{\alpha_{n,i}}{(\boldsymbol{\mu}^T\boldsymbol{\lambda}_n)^2  }$, 
where $\alpha_{n,i} = e^{j\frac{2\pi i n}{N}}$ for $0 \leq i \leq L-1$, and $\alpha_{n,i} = je^{j\frac{2\pi (i-L)n}{N}}$ for $L \leq i \leq 2L-1$.
Similarly, $\frac{\partial^2 ({H}_n^{-1})}{\partial \widetilde{b}_i^2} = \frac{2}{H_n^3} \frac{\partial {H}_n}{\partial \widetilde{b}_i}|_{\widetilde{\mathbf{b}}=\mathbf{0}} = \frac{\alpha_{n,i}}{(\boldsymbol{\mu}^T\boldsymbol{\lambda}_n)^3  }$.
Substituting the derivatives in \eqref{eq:double_derivative}, 
\begin{align}\label{scnd_drvtv_re_h_n_re_h_m}
 &[\nabla^{2}f(\mathbf{0})]_i 
 = \Bigg[\frac{\Re (\alpha_{n,i} \boldsymbol{\lambda}_n^H\boldsymbol{\mu}^*)}{(\boldsymbol{\lambda}_n^H\boldsymbol{\mu}^*\boldsymbol{\mu}^T\boldsymbol{\lambda}_n)^3  } \frac{\Re (\boldsymbol{\lambda}_m^H\boldsymbol{\mu}^* )}{(\boldsymbol{\lambda}_m^H\boldsymbol{\mu}^*\boldsymbol{\mu}^T\boldsymbol{\lambda}_m)  } \nonumber \\
 &+ \frac{\Re (\alpha_{m,i} \boldsymbol{\lambda}_m^H\boldsymbol{\mu}^*)}{(\boldsymbol{\lambda}_m^H\boldsymbol{\mu}^*\boldsymbol{\mu}^T\boldsymbol{\lambda}_m)^3  } \frac{\Re (\boldsymbol{\lambda}_n^H\boldsymbol{\mu}^* )}{(\boldsymbol{\lambda}_n^H\boldsymbol{\mu}^*\boldsymbol{\mu}^T\boldsymbol{\lambda}_n)  } \nonumber \\
 &+ 2\frac{\Re (\alpha_{n,i} \boldsymbol{\lambda}_n^H\boldsymbol{\mu}^*)}{(\boldsymbol{\lambda}_n^H\boldsymbol{\mu}^*\boldsymbol{\mu}^T\boldsymbol{\lambda}_n)^2  } \frac{\Re (\alpha_{m,i} \boldsymbol{\lambda}_m^H\boldsymbol{\mu}^*)}{(\boldsymbol{\lambda}_m^H\boldsymbol{\mu}^*\boldsymbol{\mu}^T\boldsymbol{\lambda}_m)^2  } \Bigg]
\end{align}
This can be substituted in \eqref{E_re_hn_inv_hm_inv} while noting that $\sum_{i=0}^{2L-1}(\cdot)$ applies only to $\alpha_{n,i}$ and $\alpha_{m,i}$.
Finally, we can evaluate
\begin{align}
 &\sum_{i=0}^{2L-1}[\nabla^{2}f(\mathbf{0})]_i
  = \Bigg[\underbrace{\frac{\Re (\boldsymbol{\lambda}_n^H\boldsymbol{\mu}^{*}\sum_i \alpha_{n,i} \sigma_{i}^{2} )}{(\boldsymbol{\lambda}_n^H\boldsymbol{\mu}^*\boldsymbol{\mu}^T\boldsymbol{\lambda}_n)^3  } \frac{\Re(\boldsymbol{\lambda}_m^H\boldsymbol{\mu}^*)}{(\boldsymbol{\lambda}_m^H\boldsymbol{\mu}^*\boldsymbol{\mu}^T\boldsymbol{\lambda}_m)  }}_{\rank \leq 1} \nonumber \\
  &+ \underbrace{\frac{\Re (\boldsymbol{\lambda}_m^H\boldsymbol{\mu}^{*} \sum_i \alpha_{m,i} \sigma_{i}^{2} )}{(\boldsymbol{\lambda}_m^H\boldsymbol{\mu}^*\boldsymbol{\mu}^T\boldsymbol{\lambda}_m)^3  } \frac{\Re(\boldsymbol{\lambda}_n^H\boldsymbol{\mu}^*)}{(\boldsymbol{\lambda}_n^H\boldsymbol{\mu}^*\boldsymbol{\mu}^T\boldsymbol{\lambda}_n)  }}_{\rank \leq 1} \nonumber \\ 
  &+ \underbrace{2\sum\nolimits_{i=0}^{(2L-1)}\frac{\Re (\alpha_{n,i} \boldsymbol{\lambda}_n^H\boldsymbol{\mu}^*)}{(\boldsymbol{\lambda}_n^H\boldsymbol{\mu}^*\boldsymbol{\mu}^T\boldsymbol{\lambda}_n)^2  } \frac{\Re (\alpha_{m,i} \boldsymbol{\lambda}_m^H\boldsymbol{\mu}^*)}{(\boldsymbol{\lambda}_m^H\boldsymbol{\mu}^*\boldsymbol{\mu}^T\boldsymbol{\lambda}_m)^2  }}_{\rank \leq 2L} \Bigg].
\end{align}
Thus, $\mathbb{E} \left[ \Re(\mathbf{v})\Re(\mathbf{v}^{T})\right] \in \mathbb{R}^{N \times N}$ has an $\epsilon$-rank of $(2L+2)$. The same result also holds for $\mathbb{E}\left[ \Im(\mathbf{v})\Im(\mathbf{v}^{T})\right]$. Therefore, the real-valued correlation matrix $\mathbf{R}_{\Tilde{v}\Tilde{v}}$ (and hence the covariance matrix $\mathbf{K}_{\Tilde{v}\Tilde{v}}$) has an $\epsilon$-rank of $2(2L+2) = 4(L+1)$, where $\epsilon = \mathcal{O}(\sigma_0^4)$ due to the truncated Taylor series expansion.
$\blacksquare$



\end{appendices}




\bibliographystyle{IEEEtran}
\bibliography{IEEEabrv,ref}

\begin{thebibliography}{10}
\providecommand{\url}[1]{#1}
\csname url@samestyle\endcsname
\providecommand{\newblock}{\relax}
\providecommand{\bibinfo}[2]{#2}
\providecommand{\BIBentrySTDinterwordspacing}{\spaceskip=0pt\relax}
\providecommand{\BIBentryALTinterwordstretchfactor}{4}
\providecommand{\BIBentryALTinterwordspacing}{\spaceskip=\fontdimen2\font plus
\BIBentryALTinterwordstretchfactor\fontdimen3\font minus \fontdimen4\font\relax}
\providecommand{\BIBforeignlanguage}[2]{{%
\expandafter\ifx\csname l@#1\endcsname\relax
\typeout{** WARNING: IEEEtran.bst: No hyphenation pattern has been}%
\typeout{** loaded for the language `#1'. Using the pattern for}%
\typeout{** the default language instead.}%
\else
\language=\csname l@#1\endcsname
\fi
#2}}
\providecommand{\BIBdecl}{\relax}
\BIBdecl

\bibitem{Shafin2020}
R.~Shafin, L.~Liu, V.~Chandrasekhar, H.~Chen, J.~Reed, and J.~Zhang, ``Artificial {I}ntelligence-{E}nabled {C}ellular {N}etworks: {A} {C}ritical {P}ath to beyond-{5G} and {6G},'' \emph{IEEE Wireless Commun. Mag.}, vol.~27, no.~2, pp. 212--217, 2020.

\bibitem{Lukosevicius2012}
M.~Luko{\v{s}}evi{\v{c}}ius, \emph{A Practical Guide to Applying Echo State Networks}.\hskip 1em plus 0.5em minus 0.4em\relax Springer Berlin Heidelberg, 2012, pp. 659--686.

\bibitem{zhou2020_LTD}
Z.~Zhou, L.~Liu, and H.-H. Chang, ``Learning for {D}etection: {MIMO-OFDM} {S}ymbol {D}etection {T}hrough {D}ownlink {P}ilots,'' \emph{IEEE Trans. Wireless Commun.}, vol.~19, no.~6, pp. 3712--3726, 2020.

\bibitem{zhou2020rcnet}
Z.~Zhou, L.~Liu, S.~Jere, J.~Zhang, and Y.~Yi, ``{RCNet}: {I}ncorporating {S}tructural {I}nformation {I}nto {D}eep {RNN} for {O}nline {MIMO-OFDM} {S}ymbol {D}etection {W}ith {L}imited {T}raining,'' \emph{{IEEE} Trans. Wireless Commun.}, vol.~20, no.~6, pp. 3524--3537, 2021.

\bibitem{RCstruct}
J.~Xu, Z.~Zhou, L.~Li, L.~Zheng, and L.~Liu, ``{RC-Struct: A Structure-Based Neural Network Approach for MIMO-OFDM Detection},'' \emph{{IEEE} Trans. Wireless Commun.}, vol.~21, no.~9, pp. 7181--7193, 2022.

\bibitem{LiuAI2}
H.-H. Chang, H.~Song, Y.~Yi, J.~Zhang, H.~He, and L.~Liu, ``{Distributive Dynamic Spectrum Access Through Deep Reinforcement Learning: A Reservoir Computing-Based Approach},'' \emph{IEEE Internet Things J.}, vol.~6, no.~2, pp. 1938--1948, 2019.

\bibitem{Chang2020}
H.-H. Chang, L.~Liu, and Y.~Yi, ``Deep {E}cho {S}tate {Q}-network ({DEQN}) and {I}ts {A}pplication in {D}ynamic {S}pectrum {S}haring for 5{G} and {B}eyond,'' \emph{IEEE Trans. Neur. Netw. Learn. Syst.}, vol.~33, no.~3, pp. 929--939, 2022.

\bibitem{Khani2020}
M.~Khani, M.~Alizadeh, J.~Hoydis, and P.~Fleming, ``Adaptive {N}eural {S}ignal {D}etection for {M}assive {MIMO},'' \emph{{IEEE} Trans. Wireless Commun.}, vol.~19, no.~8, pp. 5635--5648, 2020.

\bibitem{DSAComparison}
H.~Mosavat-Jahromi, Y.~Li, L.~Cai, and J.~Pan, ``{Prediction and Modeling of Spectrum Occupancy for Dynamic Spectrum Access Systems},'' \emph{IEEE Trans. Cogn. Commun. Netw.}, vol.~7, no.~3, pp. 715--728, 2021.

\bibitem{Gonon2020}
L.~Gonon, L.~Grigoryeva, and J.-P. Ortega, ``{R}isk {B}ounds for {R}eservoir {C}omputing,'' \emph{Jour. Mach. Learn. Res.}, vol.~21, no. 240, pp. 1--61, 2020.

\bibitem{Jere2023WCL}
S.~Jere, R.~Safavinejad, L.~Zheng, and L.~Liu, ``{Channel Equalization Through Reservoir Computing: A Theoretical Perspective},'' \emph{{IEEE} Wireless Commun. Lett.}, vol.~12, no.~5, pp. 774--778, 2023.

\bibitem{3GPP_CDL_channel}
\emph{5G; Study on channel model for frequencies from 0.5 to 100 GHz}, 3GPP TR 38.901, Rev. 16.1.0, 2020.

\bibitem{Bollt2021}
E.~Bollt, ``On explaining the surprising success of reservoir computing forecaster of chaos? {T}he universal machine learning dynamical system with contrast to {VAR} and {DMD},'' \emph{Chaos}, vol.~31, p. 013108, 2021.

\bibitem{Murphy2012}
K.~P. Murphy, \emph{Machine Learning: A Probabilistic Perspective}.\hskip 1em plus 0.5em minus 0.4em\relax The MIT Press, 2012.

\end{thebibliography}

\end{document}